# Hierarchical Graph Convolutional Network Built by Multiscale Atlases for Brain Disorder Diagnosis Using Functional Connectivity

Mianxin Liu, Han Zhang, Sr., *Member, IEEE*, Feng Shi, and Dinggang Shen, *Fellow, IEEE*

*Abstract*—**Functional connectivity network (FCN) data from functional magnetic resonance imaging (fMRI) is increasingly used for the diagnoses of brain disorders. However, state-of-the-art studies used to build the FCN using a single brain parcellation atlas at a certain spatial scale, which largely neglected functional interactions across different spatial scales in hierarchical manners. In this study, we propose a novel framework to perform multiscale FCN analysis for brain disorder diagnosis. We first use a set of well-defined multiscale atlases to compute multiscale FCNs. Then, we utilize biologically meaningful brain hierarchical relationships among the regions in multiscale atlases to perform nodal pooling across multiple spatial scales, namely "Atlas-guided Pooling". Accordingly, we propose a Multiscale-Atlases-based Hierarchical Graph Convolutional Network (MAHGCN), built on the stacked layers of graph convolution and the atlas-guided pooling, for a comprehensive extraction of diagnostic information from multiscale FCNs. Experiments on neuroimaging data from 1792 subjects demonstrate the effectiveness of our proposed method in the diagnoses of Alzheimer's disease (AD), the prodromal stage of AD (i.e., mild cognitive impairment [MCI]), as well as autism spectrum disorder (ASD), with accuracy of 88.9%, 78.6%, and 72.7% respectively. All results show significant advantages of our proposed method over other competing methods. This study not only demonstrates the feasibility of brain disorder diagnosis using resting-state fMRI empowered by deep learning, but also highlights that the functional interactions in the multiscale brain hierarchy are worth being explored and integrated into deep learning network architectures for better understanding the neuropathology of brain disorders.**

## I. INTRODUCTION

BRAIN disorders, such as Alzheimer's disease (AD) and autism spectrum disorder (ASD) propose a severe challenge to the global public health. The AD could induce first mild cognitive impairment (MCI) and eventually dementia in older patients. And, ASD typically affects children's social interaction and communication abilities. As of 2020, 55 million people are living with dementia, and over 75 million people are affected by ASD world-wide according to the statistics of the World Health Organization

(https://www.who.int/news-room/fact-sheets). An accurate identification as early as possible is necessary to enable effective interventions, which improve patients' outcomes. However, there is still a lack of clear understanding of the pathology of these brain disorders and well-established methods for the diagnosis.

Recent work focuses on diagnosing the human brain disorders by imaging the brain dynamics with *in vivo* brain functional imaging techniques. Among many of these techniques, resting-state functional magnetic resonance imaging (rs-fMRI) is a popular neuroimaging method that can probe spontaneous neural activities by measuring fluctuations in blood-oxygen-level-dependent (BOLD) signals. In the rs-fMRI studies, functional connectivity (FC) is the most prevalent measurement to quantitatively characterize spatiotemporal covariation patterns in the brain. Typical FC readings measure the temporal synchrony of neural activity across spatially distant regions by using pairwise Pearson correlations. The FC characterizations among all pairs of regions further lead to a whole-brain FC network (FCN). Many other FCN-constructing methods, such as the 3D maps generated by seed-based correlations or independent component analyses (ICA), can also detect neural synchronizations. FCN was found to be sensitive to capture functional changes induced by brain disorders, and has been used in individualized diagnosis for various brain disorders (see systematic reviews in [1]).

In FCN-based diagnostic studies, a long-lasting debate is about how to efficiently extract disorder-related features from the FCN. Conventional studies typically reshaped FCNs into vectors and then inputted them to classical machine-learning based classifiers, such as a support vector machine (SVM) [2]–[4]. However, such vectorizations may lose the disorder-related information carried on the topological structures of FCNs. Attributing to a rapid development of the deep learning technology, graph convolution neural networks (GCNs) allow implementations of artificial neural network algorithms on graph structures, thus providing suitable representations for the disorder-related information of FCNs. Briefly, a typical GCN deals with individual FCN and integrates nodal features along

This work was supported in part by National Natural Science Foundation of China (grant number 62131015), Science and Technology Commission of Shanghai Municipality (STCSM) (grant number 21010502600), and the National Key Scientific Instrument Development Program (No. 82027808).

Mianxin Liu and Han Zhang are with School of Biomedical Engineering, ShanghaiTech University, Shanghai 201210, China (email: liumx1@shanghaitech.edu.cn; zhanghan2@shanghaitech.edu.cn)

Feng Shi is with Department of Research and Development, Shanghai United Imaging Intelligence Co., Ltd., Shanghai 200232, China (email: feng.shi@uii-ai.com)

Dinggang Shen is with School of Biomedical Engineering, ShanghaiTech University, Shanghai 201210, China, and also with Department of Research and Development, Shanghai United Imaging Intelligence Co., Ltd., Shanghai 200232, China (Dinggang.Shen@gmail.com)

with the graph-topology neighbors based on the graph Laplacian and generates the representation for the entire FCN, which can be further used in identifications of brain disorders. Notably, another stream to apply GCN is to regard each individual data as a node and inter-individual similarity as a link. They perform nodal classification in the population-based graph to achieve individual diagnosis, which we call as "nodal classification GCN (nc-GCN)". The GCN-based brain disorder diagnosis has earned great attention, in diagnosing ASD [5], [6], major depressive disorder [7], [8], MCI, and AD [9]–[11].

However, so far, most of the existing studies, including those with GCNs, arguably ignored crucial multiscale architectures of the brain functional networks [12], [13]. Actually, in the brain, the nodes at multiple spatial scales are often organized in a hierarchical modular structure, e.g., individual neurons often connect to each other to form a "neural mass", many of which organize as neural circuits at a larger scale; and then neural circuits of many functional subsystems further form large-scale functional networks through inter-subnetwork connectivities [14], [15]. Such multiscale hierarchical brain connectomes could provide a much richer knowledge about the high-level brain functioning and therefore become more meaningful in characterizing brain disorders. First, it has been found that healthy brain connectomes at different scales exhibit different topological attributes, such as degree distribution [16], centrality [13] and small-worldness [17]. Second, brain disorders could modify FCN topology at a specific or multiple scale(s). A recent study specifically investigated modular structures of the lifespan FCNs at different spatial scales [18], indicating that the degree of segregation among network modules exhibits an interaction between age and spatial scales. Third, the relationship or interaction "across" scales could also be affected by atypical brain states, as revealed by a recent study on "cross-scale" functional interactions (namely "associated high-order FC") under the brain disorders [19].

Indeed, the latest studies started to focus on mining multiscale information from FCNs in data-driven manners [20], [21]. The advanced GCN studies have attempted to extract multiscale hierarchical topological features from the given graph, by learning various pooling operations on the nodes. Ying *et al.* proposed a coarse-graining based pooling "DIFFPOOL" and built a hierarchical GCN [22]. It learns the "cluster assignments" for nodes to aggregate the nodes into a set of clusters and to update the connectivity as well. The coarsened nodes and connectivities are forwarded to the further repeated GCN and DIFFPOOL layers, which can thus estimate multiscale community structures of the graph. Based on the DIFFPOOL, Xing *et al.* generated multiscale graph representations from FCNs and detected the diagnostic brain region clusters in MCI and obsessive-compulsive disorder patients during automatic diagnoses [20]. Also, Gao *et al.* proposed a "gPool" operation in a "Graph-U-Net" to generate a hierarchical representation of a given graph [23]. gPOOL learns a hierarchical down-sampling scheme from a single-scale graph based on the "importance" of nodal features, learned via learnable linear projections. It thus gradually focuses on a small set of nodes and their corresponding sub-networks. Following gPOOL, "self-attention graph pooling (SAGPOOL)" is further proposed to refine the hierarchical down-sampling by learning the nodal importance via GCNs on the graph [24]. Li *et al* adopted the idea of "gPool" and applied it to build hierarchical GCN for ASD detection [21].

However, all the above-mentioned studies can only mine multiscale features from a single-scale FCN, built by a predefined parcellation scheme of the atlas. On the other hand, multiscale brain atlases have been increasingly studied, which naturally provides FCNs at different spatial scales and brain hierarchies with more solid neuroscience knowledge. Recently, Yao *et al.* proposed a mutual multi-scale triplet GCN (MMTGCN) for brain disorder classifications based on the multiscale atlas information and the inter-subject sample relationship [25]. In addition to the "triplet sampling" strategy to learn inter-subject relationship, they further proposed a "mutual learning strategy" to collaboratively train multiscale encoders to generate aligned and optimal representations of the individual brain networks.

Lastly, the advanced brain atlas study offers refined multiscale parcellation and the biologically-validated prior brain hierarchy as spatial relationship among multiscale atlases [26]. In this work, we propose a novel multiscale-atlases-based hierarchical GCN architecture, to respect and utilize the prior brain network hierarchy under the GCN framework and to more effectively and sensitively capture the disorder-related intra- and inter-scale topological changes in FCNs. Specifically, we first implement multiscale atlases to construct multiscale FCNs for avoiding the potential bias in single-scale analysis and to comprehensively characterize the brain network. We also invent an Atlas-guided Pooling (AP) to perform nodal aggregation based on the prior hierarchical relationship defined in the multiscale atlases. Specifically, in each stage of nodal pooling, we update the corresponding graph with the empirical FCNs from multiscale atlases. We thus built a hierarchical architecture by stacking the GCNs and AP layers to extract multiscale disorder-related features for final diagnoses. We utilize three open-access datasets and perform diagnoses of AD, MCI, and ASD to demonstrate the effectiveness of our proposed method.

The main contributions of the presented study are three-fold. First, to the best of our knowledge, our method is the first attempt to embed the guiding priors from multiscale brain atlases into the hierarchical graph convolutional neural networks for brain disorder diagnoses. Our method distinguishes itself from other GCN methods as a novel and biologically-meaningful framework to naturally integrate domain knowledge into a data-driven learning-based method. Second, experiments demonstrate that our method can effectively detect more disorder-related features than other single-scale-based methods (including data-driven graph pooling) and other multiscale-atlases-based methods with classical fusion schemes, and thus promote precise brain disorder diagnoses. Third, our method contributes to the improvement of interpretability for artificial neural networks and also revealing of pathological factors of brain disorder among multiscale brain connectomes.

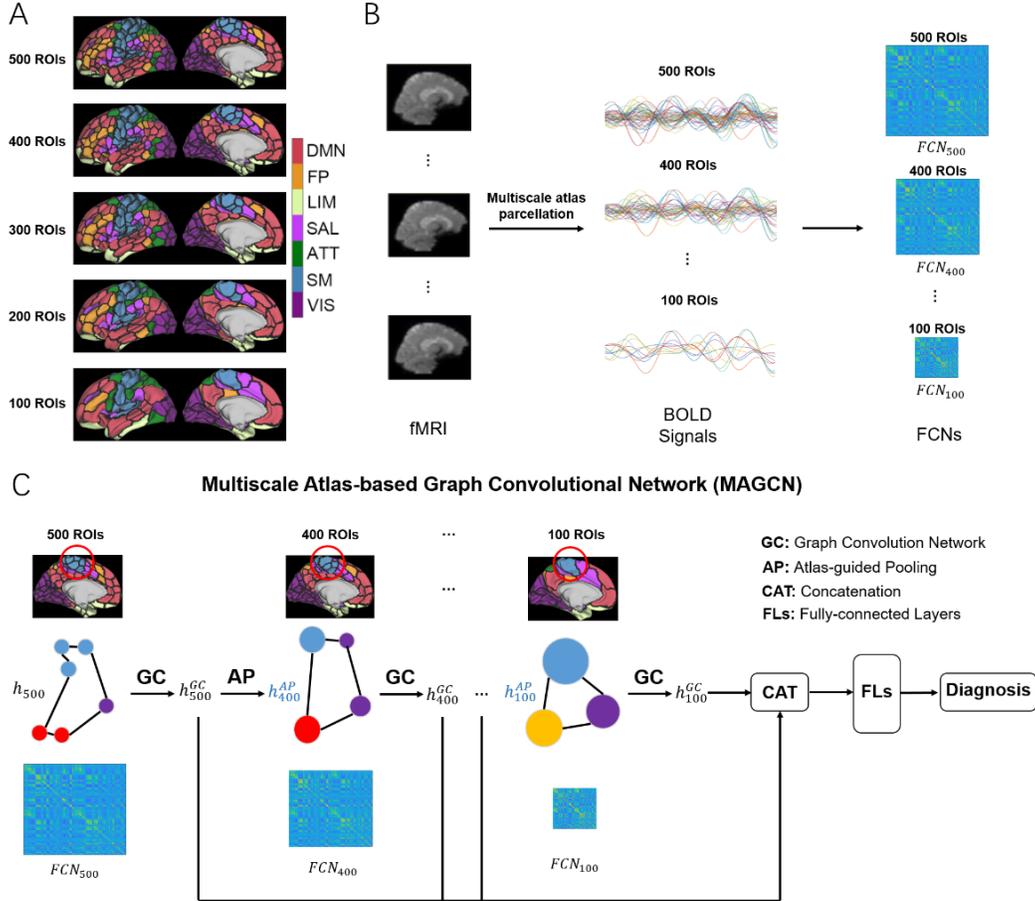

Fig. 1. Schematic illustration of the proposed Multiscale-Atlas-based hierarchy GCN (MAHGCN) workflow. For each individual, a set of predefined multiscale atlases (A) are applied to the fMRI to obtain regional averaged BOLD signals and build FCNs at different scales (B). The colors in (A) encode different resting-state networks (RSNs). The names of RSNs are attached aside the colorbar, including default mode network (DMN), frontoparietal network (FP), limbic network (LIM), salience network (SAL), attention network (ATT), somatomotor network (SM), and visual network (VIS). (C). MAHGCN extracts information from the multiscale FCNs, based on the atlas-guided pooling using hierarchical relationship between neighboring-scale atlases. The extracted features from each scale are integrated by concatenation and further processed to generate an individualized diagnosis.

A part of this work was reported at MICCAI 2021 conference [27]. In this paper, we have added new variants of the proposed method by generalizing its pooling operations. We have also added new analyses to demonstrate feature interpretability. In addition, we have included more detailed descriptions of method implementations, systematic testing of effects of parameter changing, as well as extensive experiments on additional datasets to evaluate our proposed method.

## II. MATERIALS AND METHODS

### A. Subjects and rs-fMRI data preprocessing

We use the rs-fMRI data from three datasets: i) Alzheimer's Disease Neuroimaging Initiative (ADNI-2 and ADNI-GO; http://adni.loni.usc.edu/) [28]; ii) Open Access Series of Imaging Studies (OASIS-3; https://www.oasis-brains.org/) [29]; and iii) Autism Brain Imaging Data Exchange (ABIDE-I;

https://fcon_1000.projects.nitrc.org/indi/abide/) [30]. Table 1 gives the overall information of subjects in the three datasets. Chi-square test and two-sample t-test are used to confirm no significant discrepancies in gender and age across groups (i.e., p-value>0.05), while there are significant differences in MMSE between HC and AD subjects. Specifically, the ADNI dataset is used as the main dataset for illustrations, while the OASIS and ABIDE datasets serve as additional independent datasets to partially evaluate the generalizability of our proposed method.

TABLE I
CHARACTERISTICS OF EACH STUDY'S SUBJECTS

| Study | Group | Gender (M/F) | Age | MMSE |
|-------|-------|--------------|-----|------|
| ADNI | HC | 148/216 | 71.8±6.3 | 29.1±1.1 |
| | eMCI | 56/61 | 70.5±6.8 | 28.3±1.7 |
| | AD | 25/25 | 72.8±8.4 | 22.9±2.5 |
| OASIS | HC | 107/100 | 72.3±4.5 | 28.9±1.4 |
| | AD | 28/15 | 73.2±6.6 | 25.6±3.4 |
| ABIDE | HC | 434/78 | 17.6±7.7 | \ |
| | ASD | 436/63 | 17.2±8.5 | \ |

For the ADNI dataset, a totally of 531 fMRI data from different subjects at baseline scans are selected (with no repeated scan for any single subject, which is same for the OASIS and ABIDE data), which contains 364 normal control (NC), 117 early MCI (eMCI), and 50 AD subjects. In ADNI, eMCI is defined to describe the subjects exhibiting cognitive decline but subtler than those at MCI, measured by the Wechsler Memory Scale neuropsychological test. Accordingly, the eMCI state can be more difficult to detect based on FCN. Each fMRI data is acquired with TR=3s, TE=30ms, Flip angle=80°, resolution=3.3×3.3×3.3mm³, 48 axial slices, and 420s in duration (140 volumes).

In the OASIS dataset, based on the availability, we include 207 NC and 43 AD subjects in the study. The fMRI acquisition protocols are TR=2.2s, TE=27ms, resolution=4×4×4mm³, 36 axial slices, and 372s in duration (169 volumes).

From the ABIDE datasets, we select scans with a duration longer than 300s, yielding 512 NC and 499 ASD subjects. As a multi-site dataset, the acquisition protocols and diagnostic criteria in ABIDE are varying according to data collection sites [30]. Overall, the scanning parameters are: TR=1.5-3s, TE=15-33ms, in-plane resolution $3 \times 3$-$3.438 \times 3.438$mm², slice thickness=3-4.5mm, 28-40 axial slices, and 304-486s in duration (120-300 volumes).

We applied well-accepted toolboxes, AFNI [31] (for ADNI) and DPARSF [32] (for ABIDE and OASIS), to perform a standardized preprocessing procedure for fMRI data. In particular, the first 5 or 10 volumes of each image are discarded due to potential non-equilibrium magnetization. The rigid-body transformation is performed to correct the subject's head motion. There is no large head motion (i.e., larger than 2mm or 2°) found in the used subjects. We do not further perform scrubbing/censoring of data as it may introduce additional artifacts [33]. The signals of white matter, cerebrospinal fluid, and head motion are regarded as nuisance covariates, and are regressed out from individual data. The fMRI images are then normalized to the Montreal Neurological Institute (MNI) space and spatially smoothed with a Gaussian kernel with full width at half maximum (FWHM) of 6×6×6mm³. The BOLD signals are further band-pass filtered ($0.01 \leq f \leq 0.1$ Hz) to remove the neural-irrelevant high-frequency noises and low-frequency drift from MRI. For ABIDE dataset, since volumes of scans are different among the collecting sites, we use its minimum common length, i.e., 115 volumes, from the middle of the preprocessed fMRI sequence for further processing.

### B. Multiscale atlases for brain parcellation and FCN construction

Schafer *et al.* provide a set of atlases for multiscale brain parcellation [26], which is used in this paper for generating multiscale FCNs and guiding the node pooling across scales. The atlases are generated by clustering voxels (or vertices) based on FC patterns by considering both global similarity and spatial proximity. Using different resolution parameters during the clustering results in a set of brain functional parcellation at multiple scales, ranging from 100 to 1000 regions of interest (ROIs). These ROIs belong to seven resting-state functional networks (RSNs) defined in [34]. It can be observed that the

RSN structures are largely preserved after parcellations at all scales (Fig. 1A). The relationship among the atlases at different scales thus characterizes a biologically-meaningful functional hierarchy.

Given an atlas at a specific spatial scale, the ROI-level signals can be extracted by averaging voxel-level BOLD signals within each ROI. The FCN at the given scale $R$ is then computed by Pearson correlation among all pairs of ROI-level signals and denoted as $FCN_R$ (Fig. 1B). To demonstrate our method and reduce computational and storage costs, in this study, we use the first five scales, i.e., from 100 to 500 ROIs, respectively (note that this method can be easily extended to other scales). Therefore, $R$ has a range from 100 to 500. The effect of increasing and decreasing the range of scales to be integrated is explored in later analyses, which can be extended to infer the situations when more scales are included.

### C. Multiscale-Atlases-based Hierachical Graph Convolutional Network (MAHGCN)

#### 1) Graph convolutional network

We implement the spectral graph convolution [35] to build GCN, which applies a series of approximation to obtain a simple formula of graph convolution operation and is the most popular version of the GCN. For a given graph with adjacency matrix $A$ with $R$ nodes, its corresponding symmetric normalized Laplacian $L$ is

$$L = I - D^{-\frac{1}{2}} A D^{-\frac{1}{2}}, \quad (1)$$

where $I$ is the identity matrix and $D$ is the degree matrix of $A$, yielded by filling the diagonal elements with the degree ($\sum_i A_{ij}$) of $A$ and the off-diagonal elements with 0. The eigenvectors $V$ of graph Laplacian defines the Fourier basis of the graph. With $V$, the graph convolution operation on the inputted feature vector $h$ (with dimensions $R \times ch^{in}$, where $ch^{in}$ refers to the dimension/channel size in the inputted node feature) can be defined in the spectral domain,

$$g_\theta * h = V\big((V^T g_\theta) \times (V^T h)\big), \quad (2)$$

where $g_\theta$ denotes the graph convolutional kernel with learnable parameters $\theta$, and $V^T h$ is the Fourier transform of feature vector $h$.

Direct estimation of $g_\theta$ can be computationally expensive. According to [35], we apply a simplified operation where the outputted feature from GCN $h^{GC}$ ($R \times ch^{out}$) is given by

$$h^{GC} = \sigma(\tilde{D}^{-\frac{1}{2}} \tilde{A} \tilde{D}^{-\frac{1}{2}} h\theta). \quad (3)$$

$\tilde{A} = A + I$, and $\tilde{D}$ is the corresponding degree matrix of $\tilde{A}$. $\sigma$ is a ReLU activation function. The $ch^{out}$ depends on the configuration of kernel $\theta$ in GCN. In this paper, $ch^{out}$ is set to 1 for all built GCNs.

#### 2) Atlas-guided pooling

We further define the "atlas-guided pooling (AP)" to bridge the GCNs at different scales and guide the fine-to-coarse processing of graph features. Multiple small ROIs in fine-scale parcellation could be merged into one ROI in coarse parcellation. The spatial overlapping of ROIs defined at different scales is thus utilized to approximately characterize the inter-scale hierarchy. We construct the mapping matrix between atlases by checking spatial overlapping between ROIs defined at different scales. Formally, from the atlas $\mathcal{R}$ (with $R$ ROIs/nodes) to the atlas $\wp$ (with $P$ ROIs/nodes, where $R > P$), the atlas mapping matrix $M_{\mathcal{R}\to\wp}$ (a $R \times P$ matrix) is defined as follows. First, we compute how many voxels of ROI $i$ in the finer-scale atlas $\mathcal{R}$ are spatially overlapping with ROI $j$ in the coarser-scale atlas $\wp$, and further calculate the overlapping ratio $\rho$ between the number of overlapping voxels in ROI $i$ and the total number of voxels in ROI $i$. Second, a threshold $Th$ is applied to the overlapping ratio $\rho$ to define the inter-scale mapping matrix in the AP:

$$M_{\mathcal{R}\to\wp}(i,j) = \begin{cases} 1, & \rho > Th \\ 0, & \text{Otherwise} \end{cases}. \quad (4)$$

The mapping matrix $M_{\mathcal{R}\to\wp}$ aims to pool the nodal features defined in the finer-scale atlas into the nodal features in the coarser-scale atlas. The GCN-outputted feature vector $h_R^{GC}$ defined in the atlas $\mathcal{R}$ can be mapped to $h_P^{AP}$ for the atlas $\wp$ by a simple matrix multiplication with the transported mapping matrix, $(M_{\mathcal{R}\to\wp})^T h_R^{GC} = h_P^{AP}$.

Note that the above-defined multiplication-based mapping performs a sum pooling on the nodal features across scales from the mathematical viewpoint. One can further define various pooling operations, such as average pooling and max pooling, with a slight modification.

### 3) Stacks of GCNs and APs for hierarchical architecture

After defining the AP, we can then build the multiscale-atlases-based hierarchical graph convolutional network (MAHGCN) with stacks of GCNs and APs in a hierarchical manner. In this paper, the MHAGCN integrating all five scales will be used as a full method for demonstrations. We extract FCNs from five scales (100, 200, ..., 500 ROIs) at maximum, which can be used to build a five-layer MAHGCN (as illustrated in Fig. 1C). Different scale ranges are also tested in later experiments). In particular, MAHGCN starts with a graph convolution between the nodal features and the FCN from the finest parcellation (i.e., the 500-ROI atlas in the example of Fig. 1C). As the initial inputted nodal features, a one-hot feature is used here, i.e., a $500 \times 500$ identity matrix, rather than the features extracted from brain data. This is because the effect of choosing different nodal features is not in the scope of our study. And, such a null configuration (one-hot feature as used in [35]) will make the tested models focus on topological information in FCNs and thus provide an extendable baseline. Using the AP, the outputted feature vector $h_{500}^{GC}$ (a $500 \times 1$ vector) is then converted to $h_{400}^{AP}$ (a $400 \times 1$ vector), which is further used in the next-layer graph convolution with $FCN_{400}$. The hierarchical graph convolution continues until FCN from the coarsest-scale atlas (i.e., the 100-ROI atlas) is processed.

A skip connection design, inspired by [36], is also implemented to allow outputs from all the scales to be involved in the final prediction. The outputs from all the scales are concatenated to form a fused representation for the state of the multiscale brain system and sent to two fully-connected layers (FLs). In Fig. 1C, the fused representation is $h_{1500}^{GC}$, with its size depending on the range of integrated scale (i.e., $1500 \times 1$ at most, and $300 \times 1$ at least). The first FL compresses the fused representation from MAHGCN into a $64 \times 1$ vector (which is fixed for variable ranges of integrated scale) and the second one converts the outputs from the first FL into the predicted probability for the two classes with a Softmax function (a $2 \times 1$ vector). During the implementation, the GCNs are associated with dropout (rate = 0.3), and the FLs are with the batch normalization and ReLU activation functions. These settings for GCN and FLs are kept in other methods for comparison.

### D. Model explainability analysis

To examine the explainability of our proposed model, we apply the Gradient-guided Class Activation Map (Grad-CAM) algorithm [37] to MAHGCN. By computing the Grad-CAM from GCNs at different scales, the important brain regions, defined by multiscale atlases during the predictions, can be discovered.

Briefly, the Grad-CAM tracks the gradient of the prediction output for a certain class (i.e., the normal or disorder state) with respect to each element in the feature maps outputted from the intermediate layers in the neural network (i.e., the $h_R^{GC}$). The gradient can be interpreted as the amount of influence of the intermediate feature elements on the outputted decision-making, and thus can be regarded as the importance of each feature element. The class activation map (CAM) is computed by the product between gradient map and feature map at individual data. Since the elements that positively contribute to the final prediction are more focused, the negative part of the individual CAM is further filtered out with a ReLU function.

## III. Experiments

### A. Experimental settings

#### 1) Implementation

The neural network based models are implemented using the open-source framework Pytorch [38] in Python (Version 3.6.2). The training of the network is accelerated by one Nvidia GTX 3080 GPU. To address the imbalance between positive and negative sample sizes, we apply weighted cross-entropy as a loss function, where the weights are set as the inverse of the sample ratio in the training set. Each model is trained with epoch=100, learning rate=0.001, and batch size=30. The time cost for training MAHGCN is around 30 minutes for ADNI and OASIS while 70 minutes for ABIDE. The parameters in each model are initialized with random weights. The Adam [39] is used as an optimizer, with weight decay=0.01 for avoiding overfitting.

#### 2) Validation scheme and evaluation metrics

Five times of holdout cross-validation are carried out to evaluate the models. The dataset is randomly assigned into the training and testing sets with 80-20 split for five times. And then each model is trained and tested in the five different training-testing assignments, respectively. The accuracy (ACC), sensitivity (SEN), specificity (SPE), and area under the receiver operating-characteristic curve (Area Under ROC, AUC) are chosen as performance metrics. Since the sample sizes of different classes are imbalanced in ADNI and OASIS datasets, AUC can be a more informative assessment for evaluating performance there. The above-mentioned performance metrics from each method are generally provided as the mean and standard deviation over five times of cross-validation. Where possible, we report the confidence interval (CI) for more comprehensive statistical information.

### 3) Statistical analysis

To demonstrate the significance of improvement by our proposed method, we conduct statistical comparisons for performance metrics from cross-validations, with both one-sided paired t-tests and Wilcoxon signed-rank test to examine the significance. The Wilcoxon signed-rank test is a non-parametric test on the data median, and has no assumption on the normality of the data distribution. Both types of tests are performed based on the build-in functions in Matlab, i.e., *ttest* and *signrank* (R2020b, Mathworks Inc, USA).

### B. Comparison methods

First, the single-scale *GCN* method is adopted to provide a baseline to be compared with multiscale atlases based methods.

- **GCN:** The GCN is based on [35] and the configurations described in the Method parts. The GCN is single-layered, which is followed by two FLs to generate a prediction of the status for each subject. We report the results from GCNs based on FCNs at 500-ROI scales for better a comparison with the following experiments.

Second, several representative data-driven methods are included for comparison including DiffPool [22], gPool [23], and SAGpool [24]. To make fair comparison, we mimic the structure of MHAGCN with five scales and let the DiffPool (DP), gPool (GP) and SAGPool (SAGP) mine the 500-ROI FCN and generate 400-, 300-, 200-, 100-ROI FCN in the same hierarchical manner. After layers of GCN and pooling, the encoded features are concatenated and fed into two FLs, where configurations are the same as the main architecture of MAHGCN as described above.

- **DP:** The DP is akin to a clustering method, which learns a nodal assignment/clustering matrix based on the features extracted from the nodal and graph features by a GCN. The assignment matrix serves as the mapping matrix $M_{\mathcal{R} \rightarrow \wp}$ in our method, and can be used to generate the clustered/coarse-grained graph for further processing with GCNs. Specifically, they further proposed an auxiliary link prediction objective and entropy regularization to ensure the learned assignment matrix to be sparse and modularized.

TABLE 2

CLASSIFICATION RESULTS FROM ALL COMPARISON METHODS IN FORM OF "MEAN(STANDARD DEVIATION)". THE BLACK STAR (*) BY THE METRIC FROM MAHGCN INDICATE SIGNIFICANT IMPROVEMENT AT P-VALUE<0.05 LEVEL WHEN COMPARED TO THE SECOND HIGHEST PERFORMANCE, USING EITHER ONE-SIDED PAIRED T-TEST OR WILCOXON SIGNED-RANK TEST. THE DIAMOND (◊) INDICATES THE PERFORMANCE FROM MAHGCN IS SIGNIFICANTLY BETTER THAN THE ONE FROM MAGCN (I.E., MAHGCN WITHOUT AP).

| Metric | **ADNI: eMCI** | | | | | | | |
|---|---|---|---|---|---|---|---|---|
| | GCN | DP | GP | SAGP | MV | WV | MAGCN | MAHGCN |
| ACC | 0.728(0.035) | 0.753(0.018) | 0.757(0.266) | 0.732(0.031) | 0.780(0.047) | 0.777(0.049) | 0.722(0.028) | **0.786**(0.037)◊ |
| SEN | 0.559(0.022) | 0.690(0.034) | 0.682(0.021) | 0.742(0.027) | 0.678(0.166) | 0.656(0.072) | **0.762**(0.069) | 0.751(0.039) |
| SPE | 0.780(0.047) | 0.772(0.019) | 0.780(0.041) | 0.728(0.041) | 0.813(0.064) | **0.815**(0.079) | 0.710(0.038) | 0.795(0.053) |
| AUC | 0.670(0.023) | 0.731(0.022) | 0.731(0.012) | 0.735(0.021) | 0.745(0.024) | 0.735(0.028) | 0.736(0.033) | **0.773**(0.017)◊ |

| Metric | **ADNI: AD** | | | | | | | |
|---|---|---|---|---|---|---|---|---|
| | GCN | DP | GP | SAGP | MV | WV | MAGCN | MAHGCN |
| ACC | 0.839(0.042) | 0.843(0.073) | 0.858(0.037) | 0.848(0.045) | 0.884(0.045) | 0.730(0.084) | 0.843(0.055) | **0.889**(0.031) |
| SEN | 0.651(0.121) | 0.735(0.101) | 0.711(0.029) | 0.653(0.419) | 0.633(0.183) | **0.843**(0.096) | 0.680(0.097) | 0.785(0.040)◊ |
| SPE | 0.870(0.051) | 0.865(0.092) | 0.881(0.045) | 0.878(0.050) | **0.929**(0.043) | 0.708(0.116) | 0.869(0.041) | 0.907(0.041) |
| AUC | 0.760(0.050) | 0.790(0.026) | 0.796(0.010) | 0.765(0.028) | 0.781(0.089) | 0.775(0.031) | 0.775(0.038) | **0.846**(0.012)*◊ |

| Metric | **OASIS** | | | | | | | |
|---|---|---|---|---|---|---|---|---|
| | GCN | DP | GP | SAGP | MV | WV | MAGCN | MAHGCN |
| ACC | 0.704(0.080) | 0.748(0.032) | 0.760(0.108) | 0.784(0.075) | 0.720(0.087) | 0.708(0.090) | 0.752(0.084) | **0.828**(0.069) |
| SEN | 0.621(0.077) | 0.749(0.055) | 0.764(0.137) | 0.809(0.077) | 0.738(0.131) | 0.766(0.104) | 0.751(0.123) | **0.815**(0.118) |
| SPE | 0.720(0.087) | 0.740(0.115) | 0.718(0.078) | 0.653(0.063) | 0.714(0.109) | 0.694(0.120) | 0.744(0.123) | **0.829**(0.093) |
| AUC | 0.671(0.067) | 0.744(0.034) | 0.741(0.057) | 0.731(0.061) | 0.726(0.072) | 0.730(0.058) | 0.748(0.053) | **0.822**(0.041)*◊ |

| Metric | **ABIDE** | | | | | | | |
|---|---|---|---|---|---|---|---|---|
| | GCN | DP | GP | SAGP | MV | WV | MAGCN | MAHGCN |
| ACC | 0.659(0.034) | 0.676(0.016) | 0.683(0.009) | 0.677(0.007) | 0.711(0.030) | 0.681(0.032) | 0.679(0.027) | **0.727**(0.021)*◊ |
| SEN | 0.703(0.058) | 0.709(0.067) | 0.689(0.057) | 0.690(0.073) | **0.723**(0.033) | 0.681(0.050) | 0.712(0.065) | 0.698(0.029) |
| SPE | 0.615(0.044) | 0.643(0.047) | 0.677(0.052) | 0.670(0.008) | 0.697(0.055) | 0.678(0.038) | 0.650(0.100) | **0.754**(0.030)* |
| AUC | 0.659(0.032) | 0.676(0.015) | 0.683(0.009) | 0.680(0.007) | 0.710(0.030) | 0.680(0.030) | 0.681(0.025) | **0.726**(0.021) *◊ |

- **GP and SAGP:** The GP and SAGP estimated the coarse-scale graph in a down-sampling manner. In GP, the nodal features are inputted into FLs and a Softmax operation to estimate the nodal importance, based on which the graph is down-sampled into the subgraph with the top $K$ important nodes. Here, $K$ is the parameter to control the sampling rate. In SAGP, the nodal importance is learned on the nodal and topographical features by a GCN. The down-sampled graph is forwarded to the following GCNs for processing.

Third, we compare multiscale-atlases-based methods with classical fusion schemes to demonstrate the effectiveness of multiscale atlases fusion and the advantage of our proposed method, **MAHGCN**. In particular, we perform

- **MV and WV:** Decision level fusion is first implemented. A Majority Voting (MV) among predictions from respectively trained GCN-based classifiers at different scales. The prediction from each classifier is regarded as one vote, and the final predicted label is the most voted label among all the classifiers. In addition, we also perform Weighted Voting (WV), in which the predicted probability from the pre-trained GCN-based classifiers is weighted-summed with the weights learned by training.

- **MAGCN:** Moreover, feature-level fusion is performed. We train a model with multiple GCNs to independently process the inputs of multiscale FCNs. The outputs from multiple GCNs are concatenated and forwarded to two FLs. We name this model as MAGCN. Note that this model can be also viewed as **MAHGCN without AP**, and thus can also serve as an ablation study to highlight the importance of including the prior of inter-scale hierarchy among GCNs at different scales.

## IV. RESULTS

### A. Comparison study

In this experiment, MAHGCN is compared with three data-driven hierarchical graph encoders and three multiscale-atlases-based methods. The models are trained and tested in ADNI, OASIS and ABIDE datasets, respectively. In Table 2, the performances of all comparison methods on HC-*vs*-eMCI, HC-*vs*-AD and HC-*vs*-ASD classification tasks are depicted. In all classification tasks, the following consistent results can be observed. First, both hierarchical graph processing with data-driven pooling methods (DP, GP and SAGP) and the multiscale-fusion based methods outperform the single-scale based GCN, suggesting the advantage of multi-scale analyses. Second, all best performances are achieved by multiscale-atlases-based methods, rather than data-driven methods, which suggests the advantage in biological-prior based methods over data-driven methods in properly capturing multiscale brain features. Third, among all the comparison methods, our proposed method achieves the best ACC and AUC. Most improvements are showing significance. In addition, when compared with MAGCN, i.e., MAHGCN without AP, our MAHGCN provides significantly higher AUCs in all the four tasks. This demonstrates that our proposed MAHGCN based on AP can best integrate information of multiscale-atlases-based FCNs for brain disorder diagnosis.

Besides, the performance on HC-*vs*-AD classification by MAHGCN on the OASIS dataset is comparable to that yielded on ADNI dataset, which could partially support the generalizability of our proposed model on AD identification across different datasets. In addition, our method outperforms other methods in the HC-*vs*-ASD classification task using the ABIDE dataset in terms of ACC, SPE and AUC, suggesting that our proposed method is generalizable to different brain disorders other than MCI and AD.

### B. Comparison with state-of-the-art methods

TABLE 3
THE PERFORMANCES FROM THE STATE-OF-THE-ART METHODS AND OUR PROPOSED METHOD. NOTE THAT THE wck-CNN USES DYNAMICAL FCN, AND WE USE ITS RESULTS BASED ON BASELINE SCANS, WHICH ARE MORE COMPARABLE TO OUR STUDY.

| Task | Method | Subject (H/P) | Performance |
|------|--------|---------------|-------------|
| HC *vs* eMCI | MB-CNN [40] | 48/49 | ACC 0.739 |
| | wck-CNN [41] | 48/50 | ACC 0.727 |
| | MAHGCN (ours) | 364/117 | ACC 0.786 AUC 0.773 |
| HC *vs* MCI | MMTGCN [25] | 179/191 | ACC 0.860 AUC 0.903 |
| | MAHGCN (ours) | 155/148 | ACC 0.790 AUC 0.791 |
| HC *vs* AD | wck-CNN [41] | 48/50 | ACC 0.785 |
| | MAHGCN (ours) | 364/50 | ACC 0.889 AUC 0.846 |
| HC *vs* ASD | nc-GCN [5] | 468/403 | ACC 0.704 AUC 0.750 |
| | BrainGNN [21] | 43/72 | ACC 0.798 |
| | MAHGCN (ours) | 512/499 | ACC 0.727 AUC 0.726 |

We further compare performances of our method with performances from state-of-the-art studies on FCN-based diagnostic analyses. In Table 3, we group the studies based on the classification tasks, and report the number of subjects and the achieved performances. Due to the variation of sample selection, the performances may not be quantitatively comparable. In the HC-*vs*-eMCI and HC-*vs*-AD classification tasks, our method obtains better results than the advanced CNN-based methods. And, in the HC-*vs*-ASD classification task, our methods achieve slightly worse but competitive results when compared to other GCN-based methods. The possible reason could be the usage of the larger dataset that brings more difficulty to the classification task.

We further perform a comparison with the MMTGCN method [25] that also used the multiscale atlases. Particularly, in the ADNI dataset, we select 148 MCI subjects (63 males, age=74.2 $\pm$ 8.9), match them with 155 HCs (77 males, age=74.9$\pm$6.4), and apply our MAHGCN. Note that Ref. [25] also used ADNI dataset and the ratio between HC and MCI in their study is not significantly different from ours (Chi-square

test, p=0.473). Our method obtains ACC=0.790 ± 0.012 (CI=[0.773, 0.807]), SEN=0.778±0.056 (CI=[0.701, 0.856]), SPE=0.804 ± 0.046 (CI=[0.740, 0.869]), and AUC=0.791±0.007 (CI=[0.781, 0.801]), which are lower than those reported by [25] in the HC-*vs*-MCI classification. However, we emphasize that their idea and our idea are different and could be integrated together. Their idea relies more on the inter-sample relationship, which is different from our design. More specifically, MMTGCN introduces a "triplet sampling" strategy to consider the relationship among subjects at each scale and further aligns the distributions of the encoded sample among multiple scales by a "mutual learning strategy" to benefit the integration of multiscale information. Our method considers the inter-scale correlation in nodal features when hierarchically encoding the FCN from one subject. In addition, the discrepancy in the performance may also attribute to difference in the diagnostic tasks. Their strategy of using inter-sample relationship could be more beneficial for the MCI detection. Whereas, our work could be more specific to the diagnosis for the eMCI state, which is earlier than MCI and thus is more clinically significant.

### C. Ablation study

#### 1) Effects of the range of integrated scales

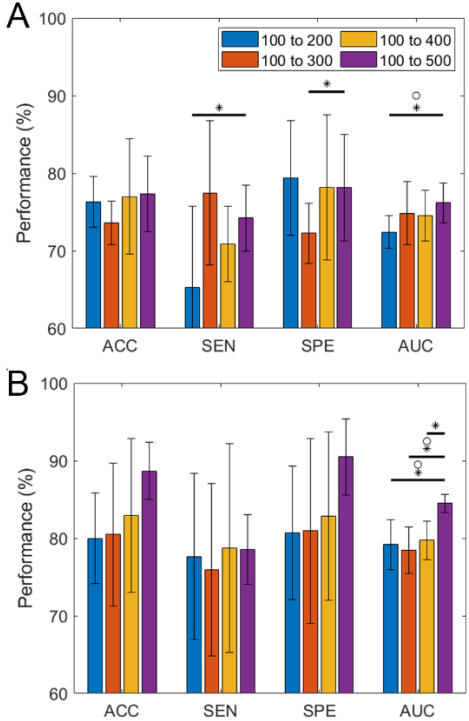

Fig. 2. Bar plots for (A) HC-*vs*-eMCI and (B) HC-*vs*-AD classification metrics using a different range of integrated scales. The black stars (*) and circles (°) indicate significance at p-value<0.05 level from the t-test and Wilcoxon signed-rank test, respectively.

We have investigated the effect of changing the range of integrated scales, using HC-*vs*-eMCI and HC-*vs*-AD classification tasks. In Fig. 2, it can be also observed that the ACCs and AUCs of MAHGCN are generally increasing with more scales included. We also try to integrate non-neighboring scales such as including 100 and 500, as well as 100, 300 and 500 in the HC-*vs*-eMCI classification, and observe similar

increasing trends. The model with 100 and 500 obtains ACC=0.714±0.037 (CI=[0.678, 0.781]), SEN=0.645±0.104 (CI=[0.500, 0.789]), SPE=0.755±0.073 (CI=[0.654, 0.857]), AUC=0.700±0.028 (CI=[0.662, 0.738]), and the model with 100, 300 and 500 yielded ACC=0.725±0.078 (CI=[0.632, 0.848]), SEN=0.674 ± 0.101 (CI=[0.534, 0.814]), SPE=0.761±0.130 (CI=[0.581, 0.942]), AUC=0.718±0.025 (CI=[0.683, 0.753]). Both observations suggest that MAHGCN can effectively integrate information from multiscale FCNs for diagnosis. However, the performance may not grow continuously with further increase of the range of integrated scales. We expect a ceiling effect could appear, although currently not being observed in our experiments. The artifacts in parcellation at fine scales could affect a meaningful hierarchical encoding. Also, more integrated scales require increased parameters in MAHGCN, leading to the risk of over-fitting on small-sample data.

#### 2) Different configurations of atlas-guided pooling

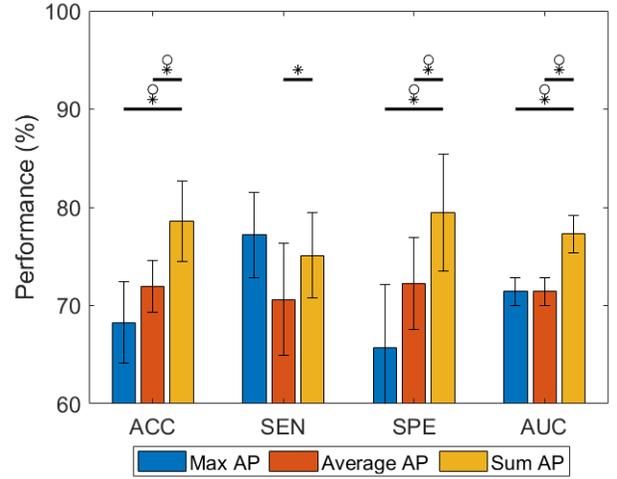

Fig. 3. Bar plots for HC-*vs*-eMCI classification metrics using different schemes of AP. HC-*vs*-eMCI classification is performed with the MAGCN integrating 100- to 500- ROI scales. The black stars (*) and circles (°) have the same meaning as in Fig.2.

We perform further ablation studies by changing the configurations of the MAHGCN. The HC-*vs*-eMCI classification, as the most difficult task, is performed to sensitively assess the changes due to modification of the model. We test the performance under different pooling schemes (i.e., using max, average, and sum pooling) in the AP. The corresponding models are called ***Max AP***, ***Average AP***, and ***Sum AP***. Note that the "Sum AP" is a default configuration we use above.

Interestingly, unlike natural image processing where the average pooling and the max pooling are the most efficient pooling schemes, we find that the sum pooling significantly outperforms other types of pooling in terms of ACC, SPE, and AUC (Fig. 3). This may attribute to the essential difference between natural image data and brain network data. Average pooling normalizes differences in a number of nodes under integration at the finer scales, and flattens differences of nodes at the coarser scales; while, the max pooling only relies on the ROI with max features to represent other ROIs at the finer scale.

These pooling schemes could ignore meaningful biological information for the brain system and thus produce decreased performance.

### 3) Effects of skip connections

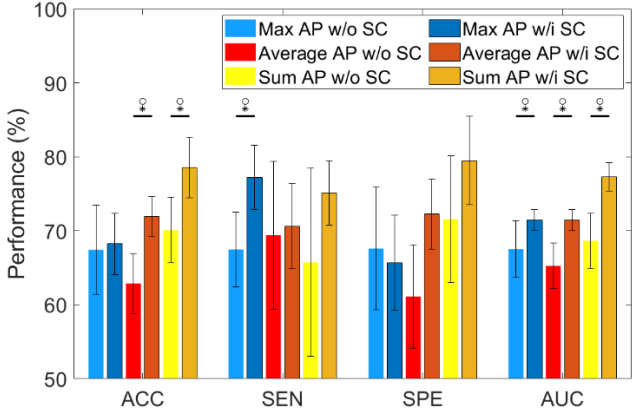

Fig 4. The bar plot for the effect of skip connections (SCs) in MAGCN with different schemes of AP. HC-*vs*-eMCI classification is performed with the MAGCN integrating 100- to 500- ROI scales. The respective performance for MAGCN without SC is in brighter color with a black box.

In Fig. 4, we systematically test MAHGCNs with different schemes of AP and their corresponding versions without SCs. The SCs significantly improve AUC in MAHGCNs with all schemes of AP. This is reasonable since the final output from MAHGCN only provides an abstract feature shared by FCNs at all scales, while SCs allow the reservation of scale-specific information at all scales. In addition, the SC could overcome the gradient vanishing and benefit to the optimization of MAHGCN.

### 4) Effects of thresholds for atlas mapping matrix

In this paper, we simply use $Th=0$, under which any sharing between the ROIs in atlases is considered as overlapping ROIs. Other settings of $Th$ are also tested, and the respective diagnosis performances are provided in Fig. 5. Results suggest that a tight threshold could diminish the hierarchical relationship among scales and degrade the performance.

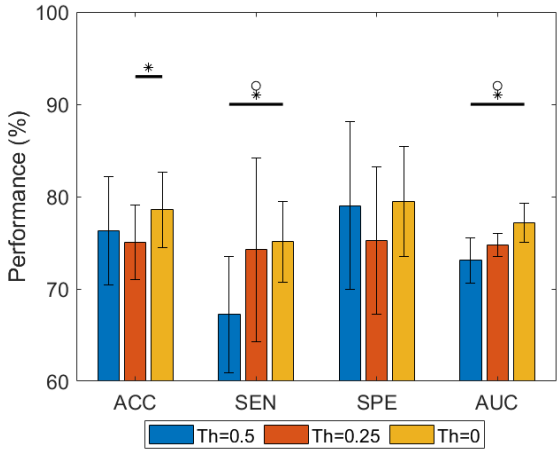

Fig.5. HC-vs-eMCI classification performances from MAHGCN models (with 100-500 ROI scales) under different overlapping thresholds ($Th$). Blue: $Th=0.5$; Red: $Th=0.25$; Yellow: $Th=0$. The black stars (*) and circles (°) have the same meaning as in Fig.2.

### 5) Effects of pre-processing pipeline

The ABIDE dataset provides pre-processed fMRI data using different pipelines, which are further quality-controlled (see https://preprocessed-connectomes-project.org/abide/). To test the effect of pre-processing on our method, we utilize those preprocessed data under two pipelines, i.e., DPARSF and Connectome Computation System (CCS, based on AFNI and FSL [42]), to investigate the effect of pre-processed pipeline on our MHAGCN. After checking data availability for both pipelines and matching the age and gender, we include 371 HC (311 males and age=16.60±6.70 years) and 325 ASD (284 males and age=17.07±8.32 years) for this analysis.

The MHAGCN achieves ACC=0.694±0.018 (CI=[0.670, 0.719]), SEN=0.709 ± 0.060 (CI=[0.625, 0.793]), SPE=0.684±0.070 (CI=[0.587, 0.781]), AUC=0.696±0.017 (CI=[0.672, 0.721]) using DPARSF, and ACC=0.706±0.016 (CI=[0.684, 0.728]), SEN=0.684±0.093 (CI=[0.589, 0.846]), SPE=0.702±0.091 (CI=[0.557, 0.810]), AUC=0.702±0.013 (CI=[0.684, 0.719]) using CCS. These two results show no significant difference using paired t-test (p=0.728) and Wilcoxon signed-rank test (p=0.526). The reduction of performance compared to results in Table 1 may attribute to the changes of sample size.

### D. Explainability analysis

In order to demonstrate the explainability of the prediction from MAHGCN, we apply Grad-CAM to explore important features during the decision making by MAHGCN. In particular, the group-wise CAMs of five models from cross-validations are first obtained by averaging the CAMs from each individual within brain disorder groups. Then, the resulting group-wise CAMs are weight-averaged according to the AUC of each model, yielding the CAMs as brain patterns (Fig. 6, A-B). In Fig. 6, C-D, we further assign class activation values in ROIs to the corresponding RSNs and compute the RSN-wise averaged class activation values.

For AD diagnosis, a pronounced finding is that the limbic network (LIM) is detected as an important predictor shared by most of the scales (100, 300, 400 and 500). This is consistent with the widely accepted understanding that AD is associated with the damage in the hippocampus related to the limbic network and thus typically exhibits amnestic cognitive impairment [43]. In addition, the default mode network (DMN) (related to executive functions) at scales of 200, 400, and 500 ROIs are found to be the second most contributive RSN, which is also consistent with the findings in the AD literature [43]. These consistencies could partially support the explainability of the features identified by our method.

For eMCI diagnoses, different scales rely on different RSNs as the most important features. Somatomotor (SM), limbic, visual (VIS), attention (ATT) and salience (SAL) networks can be identified as the predictive RSNs at different scales, and there is few RSN being consistently identified as a crucial predictor. This may also be reasonable due to complex symptoms of the eMCI involving impairments in multiple cognitive domains, including executive function, attention, and visuospatial ability or language, besides memory [44], [45].

This could deliver a new insight that eMCI damages the functioning of different cognitive domains at different scales.

## V. CONCLUSION

We have presented a novel method for brain disorder diagnosis based on information from multiscale FCNs and pre-defined multiscale atlases. Using neuroimaging data from ADNI, OASIS and ABIDE, we have demonstrated that our proposed method has a higher effectiveness in identifying brain functional deficits induced by MCI, AD, and ASD. The improvement of diagnosis by our proposed method comes from the usages of prior knowledge on brain parcellation as well as proper designs (based on brain hierarchy) to fuse multiscale spatial information in a deep-layer GCN structure. We have further investigated the features of atypical brain connectomes across different scales, and provide novel clues about the pathology of MCI and AD, paving a new path to explore multiscale-atlases-based FCN in the brain disorder diagnostic studies. Through FCN is focused in this study, we expect our method or its variations could be applied in brain network data based on other modalities [46]–[49] in the future.

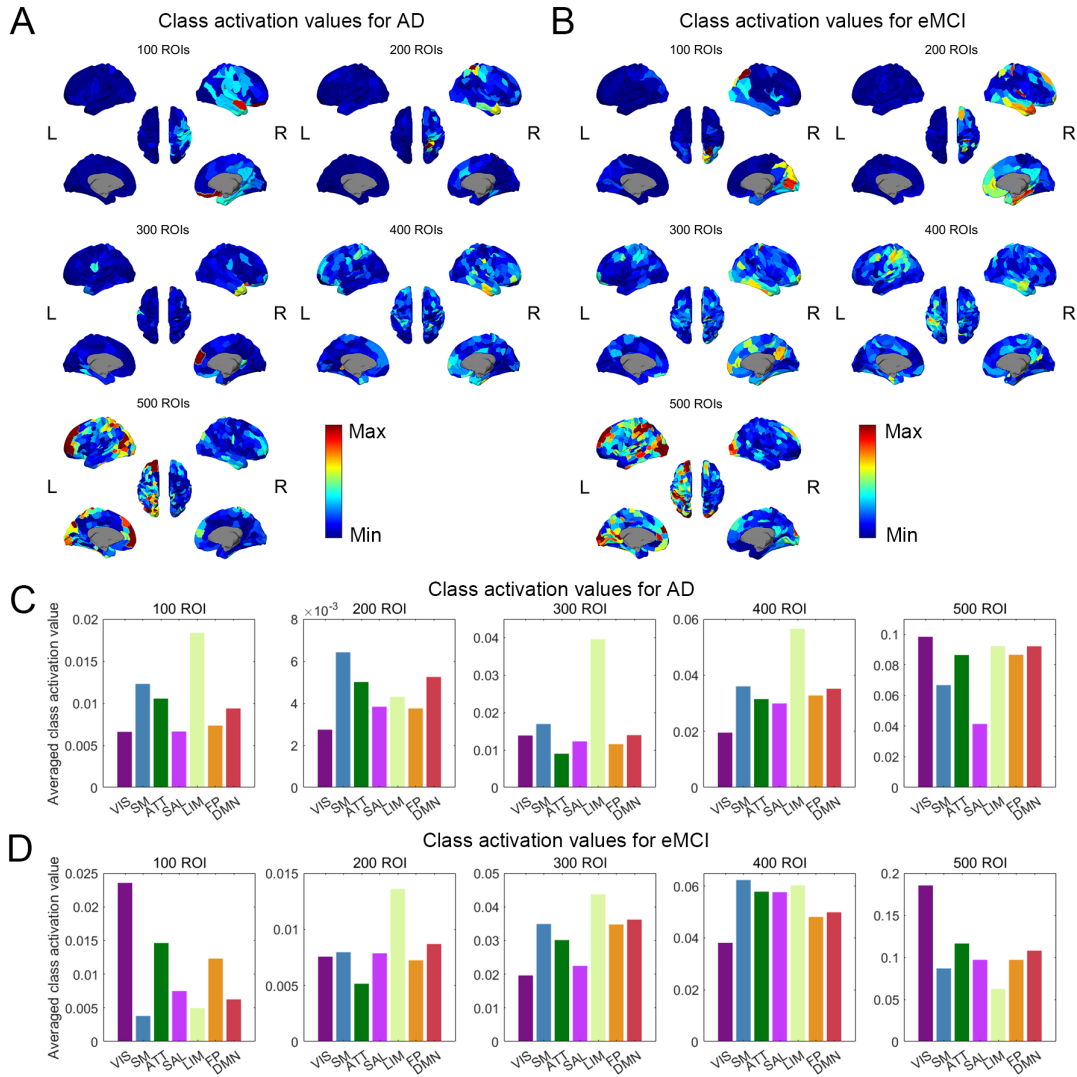

Fig 6. The RSN-wise averaged class activation values for diagnosing eMCI and AD. The color codings and short names for different RSNs are the same as Fig. 1A. (A-B) The brain patterns of class activation values for eMCI and AD; (C-D) Distributions of RSN-wise averaged class activation values for eMCI and AD.